\documentclass[%
reprint,
amsmath,amssymb,
aps,
showkeys
]{revtex4-2}

\usepackage{graphicx}
\usepackage{dcolumn}
\usepackage{bm}
\usepackage{hyperref}
\hypersetup{colorlinks,allcolors=blue}
\usepackage[table,dvipsnames]{xcolor}
\usepackage{float}
\usepackage[braket, qm]{qcircuit}
\usepackage{graphicx}
\usepackage{multirow}
\usepackage{qcircuit,amsmath}
\usepackage{longtable}
\usepackage[dvipsnames]{xcolor}

\begin{document}

\title{Comprehensive characterization of three-qubit Grover search algorithm on IBM's 127-qubit superconducting quantum computers}

\author{Muhammad AbuGhanem{$^{1}$}}
 \address{$^{1}$ Faculty of Science, Ain Shams University, Cairo, $11566$, Egypt}

\email{gaa1nem@gmail.com}

\date{\today}

\begin{abstract}

The Grover search algorithm is a pivotal advancement in quantum computing, promising a remarkable speedup over classical algorithms in searching unstructured large databases. Here, we report results for the implementation and characterization of a three-qubit Grover search algorithm using the state-of-the-art scalable quantum computing technology of superconducting quantum architectures. To delve into the algorithm's scalability and performance metrics, our investigation spans the execution of the algorithm across all eight conceivable single-result oracles, alongside nine two-result oracles, employing IBM Quantum's 127-qubit quantum computers. Moreover, we conduct five quantum state tomography experiments to precisely gauge the behavior and efficiency of our implemented algorithm under diverse conditions – ranging from noisy, noise-free environments to the complexities of real-world quantum hardware. By connecting theoretical concepts with real-world experiments, this study not only shed light on the potential of NISQ (Noisy Intermediate-Scale Quantum) computers in facilitating large-scale database searches but also offer valuable insights into the practical application of the Grover search algorithm in real-world quantum computing applications. 

\end{abstract}

\keywords{Quantum search, Grover algorithm, quantum algorithms,  applied quantum computing, quantum state tomography, IBM Quantum's superconducting quantum computers. \\
PACS: 03.67.Ac, 03.67.Dd, 03.67.Lx, 03.67.-a, 03.65.Ca, 87.55.kd, 87.55.kh}

\maketitle

\section{Introduction}

\begin{figure*} 
    \centering  
    \includegraphics[width=0.77\textwidth]{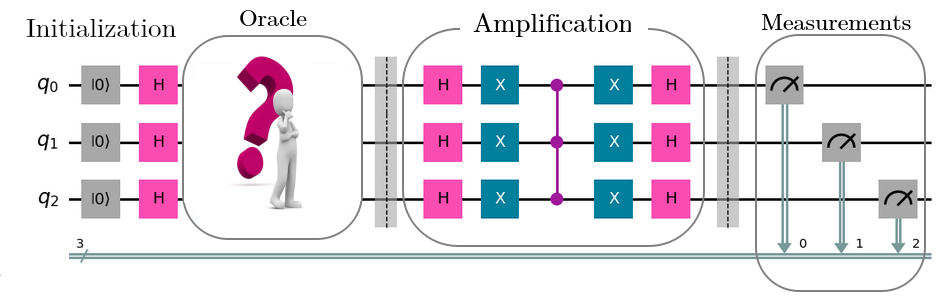}
    \vspace{-0.4cm}
\caption{A schematic circuit representation illustrating the key stages of GSA. The process includes initialization, where the qubits are prepared in a superposition state; marking, where the target item (items) in the database are identified and marked by the oracle; amplification, where the amplitude(s) of the marked item(s) are increased; and measurement, where the final result is obtained. Here, \ket{0}, H, and X represent the initialization of the qubit to the \ket{0} state, the Hadamard gate, and Pauli's X gate, respectively.}
\label{fig:GroverConcept}
    \label{fig:enter-label}
\end{figure*}

Quantum computing has emerged as a transformative field with the potential to revolutionize various domains, from cryptography and optimization to drug discovery and materials science~\citep{NISQ23}. At the heart of quantum computing lies the promise of harnessing quantum phenomena to perform computations at unprecedented speeds, surpassing the capabilities of classical computers~\citep{QAlg_Deutsch,QAlg_DJoza,QAlg_BV,QAlg_Simon,QAlg_Shor,Models19,QAlg_HHL,QAlg_VQA,QAlg_VQE,QAlg_QAOA,Light}.

Searching through extensive databases is a crucial challenge with far-reaching applications. Among the plethora of quantum algorithms developed thus far, Grover's search algorithm (GSA) stands out as a powerful technique for searching unsorted databases~\citep{Alg_Grover,Intro2}. Proposed by Lov Grover in 1996, this algorithm offers a quadratic speedup over classical algorithms, making it particularly attractive for a wide range of applications~\citep{Alg_Grover}.

Grover's algorithm occupies a central position within the realm of quantum computing, heralded for its versatility and utility across numerous disciplines. It stands as the optimal search algorithm for quantum architectures~\citep{Intro3}, finding utility as a foundational component for quantum algorithms~\citep{Intro4,Intro5}. 
Its applicability extends to 
facilitating string matching tasks~\citep{Distrib14}, 
addressing minimum search challenges~\citep{Distrib13}, 
tackling computational geometry problems~\citep{Distrib16}.
and enabling quantum dynamic programming~\citep{Distrib15}.

Successful demonstrations of searches employing two qubits have been achieved across different quantum platforms~\citep{Intro6,Intro7,Intro8,Intro9,Intro10,Intro11,Intro13}. 
However, efforts to expand these capabilities to larger search spaces have primarily been demonstrated on non-scalable NMR systems~\citep{Intro14}. Additionally, a three-qubit GSA using programmable trapped atomic ion systems was implemented in~\citep{Cfggatt17}. Reports detailing the successful execution of multi-qubit algorithms remain rare.

The motivation behind our research lies in the growing interest in practical quantum computing and the necessity to comprehend  the experimental feasibility and performance of implementing quantum algorithms on real quantum hardware, especially within the NISQ era. 
This paper presents a comprehensive study focusing on the experimental implementation and characterization of the GSA on large-scale superconducting quantum computers. Leveraging state-of-the-art quantum hardware, we aim to explore the scalability, performance, and practical challenges associated with deploying Grover's algorithm in real-world settings.

The remainder of this paper is organized as follows: 
Section~\ref{Sec:2} begins with an overview of the experimental setup and implementation details, detailing the procedures and methodologies employed to realize the GSA on the real quantum hardware. In Section~\ref{Sec:2}, we delve into the intricacies of quantum circuit design, oracle construction, and quantum state preparation, laying the groundwork for our experimental investigations. 
Subsequently, in Section~\ref{Sec:3}, 
we conduct a thorough characterization of the implemented algorithm, employing ${\cal QST}$ (quantum state tomography) experiments to assess its performance across different configurations and environmental conditions. We analyze key metrics such as algorithm success probability (${\cal ASP}$), squared statistical overlap (${\cal SSO}$), and state fidelity (${\cal F}_S$), providing insights into the algorithm's behavior and its suitability for practical applications.
Furthermore, 
Section~\ref{Sec:4} 
includes a detailed analysis and discussions, where we interpret our experimental findings, compare them with theoretical expectations, 
and explore implications for future research and development in quantum computing. 
Finally, Section~\ref{Sec:5} concludes the paper with a summary of our findings and suggestions for future research directions.

\section{Experimental setup and implementations}\label{Sec:2}

\subsection{Problem statement}

Grover’s algorithm addresses the problem of unstructured search. In this context, an unstructured search problem entails having a set of $S$ elements forming a set $\Gamma = \{\gamma_1, \gamma_2, \dots, \gamma_S\}$, along with a boolean function $f :\Gamma  \rightarrow \{0, 1\}$. The objective is to identify an element $\gamma^*$ in $\Gamma $ for which $f(\gamma^*) = 1$.

For instance, consider a scenario where we're searching for a specific phone number in a directory. The function $f(\gamma)$ evaluates whether a given phone number matches the desired one. 
The essence of the problem lies in its abstraction, such that any search task can be distilled into an evaluation of a function $f(\gamma)$, where $\gamma$ represents potential search items. Should a particular item $\gamma$ hold the solution, the function returns 1; otherwise, it returns 0. Thus, the fundamental challenge (search problem) is to uncover any such $\gamma^*$ that yields a result of 1.

Grover's algorithm embarks on this quest by tackling a classical function ($f(\gamma): \{0,1\}^n \rightarrow \{0,1\}$), where $n$ denotes the bit-size of the search space. In the classical realm, the algorithm's complexity hinges on the sheer number of times the function $f(\gamma)$ must be interrogated. In the most unfavorable scenario, this involves an exhaustive search through $S-1$ iterations, where $S=2^n$, exhaustively exploring every conceivable option. However, GSA offers a profound departure from this laborious approach, promising a remarkable quadratic acceleration. Specifically, this signifies that the algorithm can ascertain the sought-after solution with a mere ${\cal O}(\sqrt{S})$ evaluations, a stark contrast to the linear $S$ evaluations demanded by classical methods~\citep{Alg_Grover}.

\subsection{Methodology}

Grover's algorithm not only revolutionizes the speed at which search tasks are accomplished but also epitomizes the transformative potential of quantum computing in navigating complex computational landscapes with unprecedented efficiency. Its methodology comprises several pivotal stages: initialization, marking, amplification, and measurement, as depicted in Figure~\ref{fig:GroverConcept}. 
The algorithm initiates with an input state $\ket{\zeta_0} = \sum_{\chi=0}^{S-1} \ket{\chi}  \rightarrow  \ket{0}^{\otimes n}$, where it establishes a superposition of all potential search states, $ H^{\otimes n} \ket{\zeta_0} \rightarrow \ket{\zeta_1} = \frac{1}{\sqrt{S}}\sum_{\chi=0}^{S-1} \ket{\chi}$, laying the groundwork for quantum parallelism. 
Subsequently, during the marking phase, an oracle function ($\mho_f:\ket{\chi} \rightarrow (-1)^{f(\chi)} \ket{\chi}$) is employed to identify and mark the target state $\ket{\chi^*}$ or states within the superposition. 
These oracle functions alter the sign of the amplitude of the marked state(s).  Mathematically, 
for a single-search result scenario, it can be represented as $\ket{\zeta_2} = -\frac{1}{\sqrt{S}} \ket{\chi^*} + \frac{1}{\sqrt{S}} \sum_{\chi=0,\chi\neq \chi^*}^{S-1} \ket{\chi}$. 
This marking process is crucial as it enables the algorithm to concentrate computational resources on the desired outcome. 

The heart of Grover's algorithm lies in the amplification phase (see Figure~\ref{fig:GroverConcept}), where a Grover operator (${\cal GO}$) is iteratively applied to boost the probability amplitude of the marked state(s) while simultaneously diminishing the amplitudes of non-marked states. 
The ${\cal GO}$ involves reflections about the mean ($\mu_G = \frac{1}{S} \sum_{\chi=0}^{S-1} \varpi_{\chi} $) so that 
$\ket{\zeta_3} = -\varpi_{\chi^*} \ket{\chi^*} + \varpi_{\chi} \sum_{\chi=0,\chi\neq \chi^*}^{S-1} \ket{\chi}$, 
and inversions about the marked state, 
$\ket{\zeta_4} = (2 \mu_G + \varpi_{\chi^*}) \ket{\chi^*} +  (2 \mu_G - \varpi_{\chi}) \sum_{\chi=0,\chi \neq {\chi^*} }^{S-1} \ket{\chi}$, 
where $\varpi_{\chi}$ denotes the amplitude for the basis state $\ket{\chi}$. 
This iterative process is pivotal in achieving the algorithm's remarkable time complexity of ${\cal O}(\sqrt{S})$, representing a quadratic speedup over classical search algorithms. 
Finally, the algorithm concludes with a measurement stage, wherein the quantum state is measured, yielding the marked item(s) with high probability. 
The correctness of the algorithm is validated by assessing the probability of measuring the marked state.

\subsection{Design and execution of quantum oracles}

\begin{table*}
    \centering
    \caption{Single-solution GSA phase oracles and their corresponding boolean oracles, for all eight single-marked states employed in the 3-qubit GSA. Illustrated using conventional notation for quantum circuit diagrams, where $q_k$ denotes an ancillary qubit.}
   \label{tab:Oracles1}
    \begin{tabular}{c}
        \includegraphics[width=0.85\textwidth]{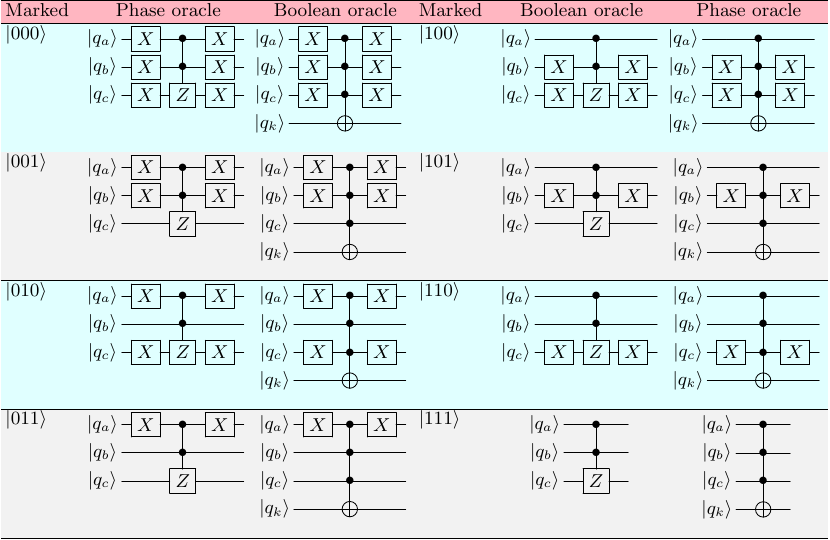} \\
    \end{tabular}
\end{table*}

\begin{table*}
    \centering
\caption{A random sample of two-solution GSA phase oracles and their corresponding boolean oracles. 
In our experimental investigation, we analyze the performance of the GSA with nine distinct 2-search phase oracles, each uniquely designed to mark two states within the search space. Illustrated using conventional notation for quantum circuit diagrams.}
\label{tab:Grover1Oracles}
    \label{tab:oracles2}
    \begin{tabular}{c}
        \includegraphics[width=0.85\textwidth]{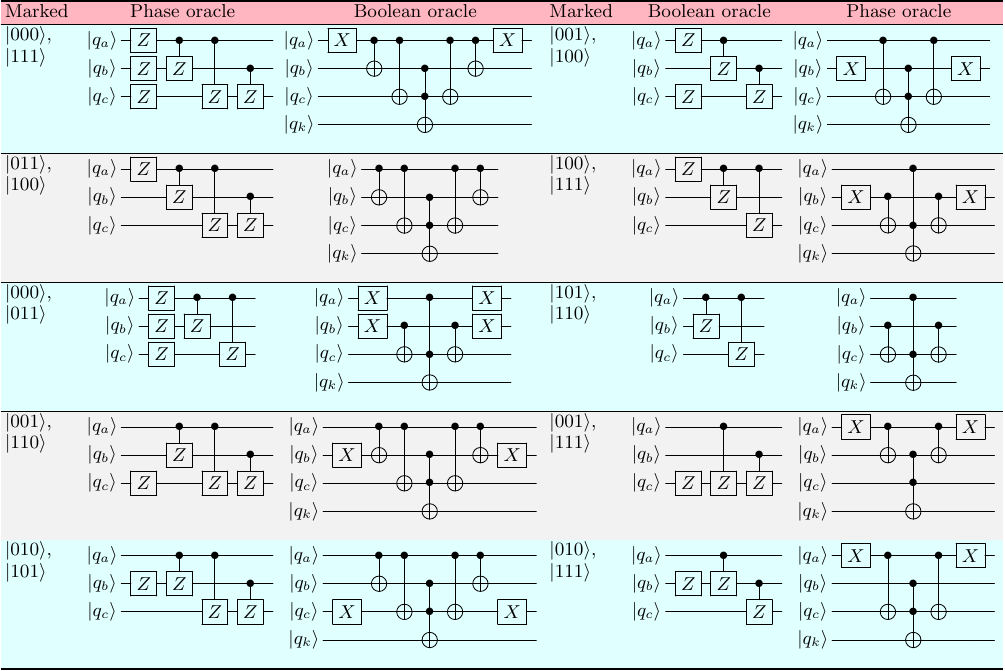} \\
    \end{tabular}
\end{table*}

Here, we implement the GSA using state-of-the-art scalable superconducting quantum computers, utilizing \( n = 3 \) qubits, which corresponds to a search database of size \( S = 2^n = 8 \). Table~\ref{tab:Oracles1} illustrates the phase oracles alongside their corresponding boolean oracles for all eight single-marked states employed in the 3-qubit GSA. These marked states are:
$\ket{000}, \, \ket{001}$, 
$\ket{010}, \, \ket{011}$, 
$\ket{100}, \, \ket{101}$, 
$\ket{110}, \, \text{and} \ket{111}$. 
Additionally, 
we scrutinize the performance of the GSA with nine distinct two-search phase oracles (as depicted in Table~\ref{tab:oracles2}), each uniquely designed to mark two states within the search space. The marked states include: 
$\ket{000}-\, \ket{111}$,
$\ket{001}-\, \ket{100}$,
$\ket{011}-\, \ket{100}$,
$\ket{010}-\, \ket{111}$,
$\ket{000}-\, \ket{110}$,
$\ket{010}-\, \ket{101}$,
$\ket{101}-\, \ket{110}$,
$\ket{101}-\, \ket{111}$, and
$\ket{100}-\, \ket{111}$. 
In our experimental setups, we employ the GSA with phase oracles, a technique previously verified in other experimental configurations. 
It is noteworthy that both phase and boolean oracles exhibit mathematical equivalence~\citep{Nilson}.

The likelihood of detecting the designated state following \( k \) iterations of the ${\cal GO}$ is a crucial metric for evaluating the algorithm's efficiency. This likelihood is articulated by the expression: 
\begin{equation}\label{P}
 {\cal P}(\ket{\chi^*})  = k \cdot \left( \left[ \frac{\pounds-2k}{\pounds} + \frac{2(\pounds-k)}{\pounds} \right] \frac{1}{\sqrt{\pounds}} \right)^2 
\end{equation}
This formulation arises from the amplitude amplification process~\citep{Intro2}, delineating the probability amplitude alterations occurring at each iteration. 

In the case of a single-solution algorithm with \( k = 1 \) iteration in GSA, the algorithmic probability of measuring the correct state after one iteration results in approximately \( 78.125\% \). This probability is significantly higher than the probability achieved by the optimal classical search strategy, which consists of a single query followed by a random guess in case of failure. In the classical strategy, the probability is calculated as:
\begin{equation}\label{PC}
{\cal P}_{\text{classical}} = \frac{k}{\pounds} + \left(1 - \frac{k}{\pounds}\right) \cdot \frac{k}{\pounds-1} 
\end{equation}
For this case, it equals $25\% $, where \( k \) represents the number of correct answers (1 in this case) and \( \pounds \) represents the total number of possible answers~\citep{Cfggatt17}. 
This comparison highlights the quantum advantage of GSA over classical search strategies, demonstrating its superior efficiency in finding solutions with fewer queries, especially in scenarios with a single correct answer.

\subsection{Results overview}

The ${\cal ASP}$ 
encapsulates the probability of successfully identifying the marked state as the conclusive result of an experiment. 
In the case of a two-solution algorithm, ${\cal ASP}$ is determined by aggregating the probabilities associated with observing each of the two marked states. 
Meanwhile, the ${\cal SSO}$ offers a quantitative measure of the extent to which observed and expected populations overlap across all states. 
The ${\cal SSO}$ 
is calculated as follows:
\begin{equation}
{\cal SSO} = \left( \sum_{m=0}^{M} \sqrt{\xi_m \cdot {\cal M}_m} \right)^2 
\end{equation}
Here, 
 \( \xi_m \) represents the expected population for each state \( m \),  while 
 \( {\cal M}_m \) signifies the measured population for each state \( m \), and 
\( M \) denotes the total number of states. 
This formulation of ${\cal SSO}$ is often used in various fields, including population studies, where it quantifies the similarity or overlap between observed and expected population distributions across different categories or states~\citep{Cfggatt30}.

In our investigation, we delved into executions of the 
GSA for both single-solution and two-solution scenarios on a 3-qubit database, exploring diverse environmental conditions such as noisy environments and leveraging real quantum computers provided by IBM Quantum~\citep{IBMQ}. 
Specifically, we employed three of IBM's 127-qubit superconducting quantum computers: \textit{ibm\_sherbrook}, \textit{ibm\_osaka}, and \textit{ibm\_kyoto}. These quantum machines are equipped with Eagle r3 quantum processors, boasting capabilities of executing a maximum of 300 circuits and 100,000 shots. 
For further insights, the key specifications and qubit characteristics of these 127-qubit quantum computers, including error rates for individual gates and readout, as well as their basis gates, are meticulously outlined in Appendix~\ref{A:1}.

While acknowledging the theoretical optimality of GSA, which suggests a runtime of ${\cal O}(\sqrt{S})$ iterations to locate the marked state within a search space of size $S$, our experimental implementations were tailored to address the practical realities of quantum hardware, while still showcasing the algorithm's effectiveness. Thus, we iterate the ${\cal GO}$ ten times for each oracle, guided by the prevalent challenges posed by errors and noise in real-world quantum hardware, particularly in the NISQ era.

\begin{figure*}
\centering
\begin{minipage}{1.0\textwidth}
\centering
\includegraphics[width=0.95\textwidth]{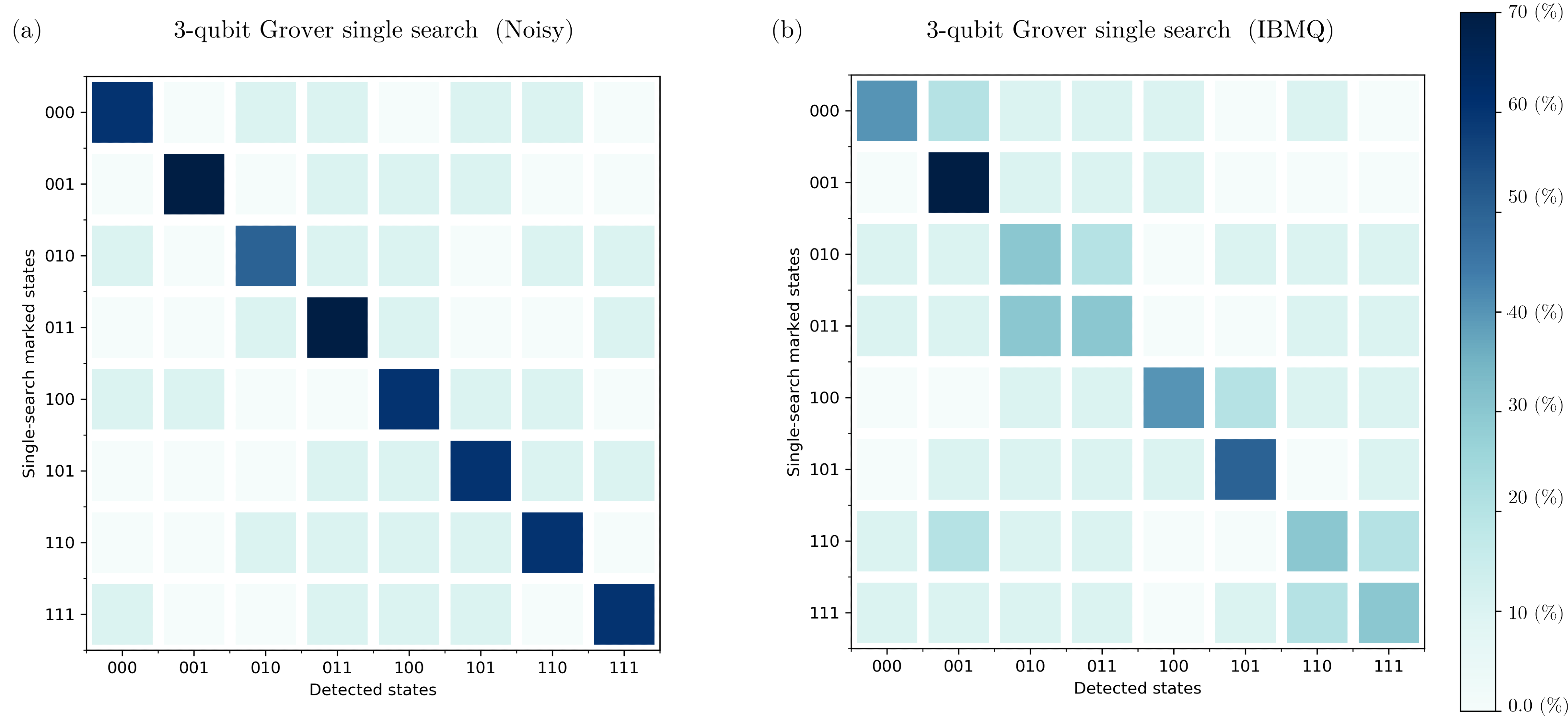}
\caption{Results obtained from executing the GSA for single-solution scenarios on a 3-qubit database (000, 001, 010, 011, 100, 101, 110, 111) in various environments. The left side presents data from algorithm execution in a noisy environment, while the right side displays data from execution on IBM Quantum's real quantum computers. The graphs illustrate the probability distribution for each output state. We observed a median ${\cal ASP}$ of 76.79\% in the noisy execution and 44.80\% on the IBM quantum computers. Additionally, we obtained a median ${\cal SSO}$ of 82.49\% in the noisy environment and 72.63\% on real IBM quantum computers. All percentages are calculated relative to the expected state $\ket{\psi_E}_\text{Single}$, defined as $\ket{\psi_E}_\text{Single} = \frac{5}{4\sqrt{2}} \ket{\chi^*} +\frac{1}{4\sqrt{2}} \sum_{\chi \neq \chi^*} \ket{\chi}$, where $\ket{\chi^*}$ represents the single marked state. The $\ket{\,}$ notation was omitted from the figures for simplicity.}
\label{fig:heat1}
\end{minipage}
\begin{minipage}{1.0\textwidth}
\centering
\includegraphics[width=0.95\textwidth]{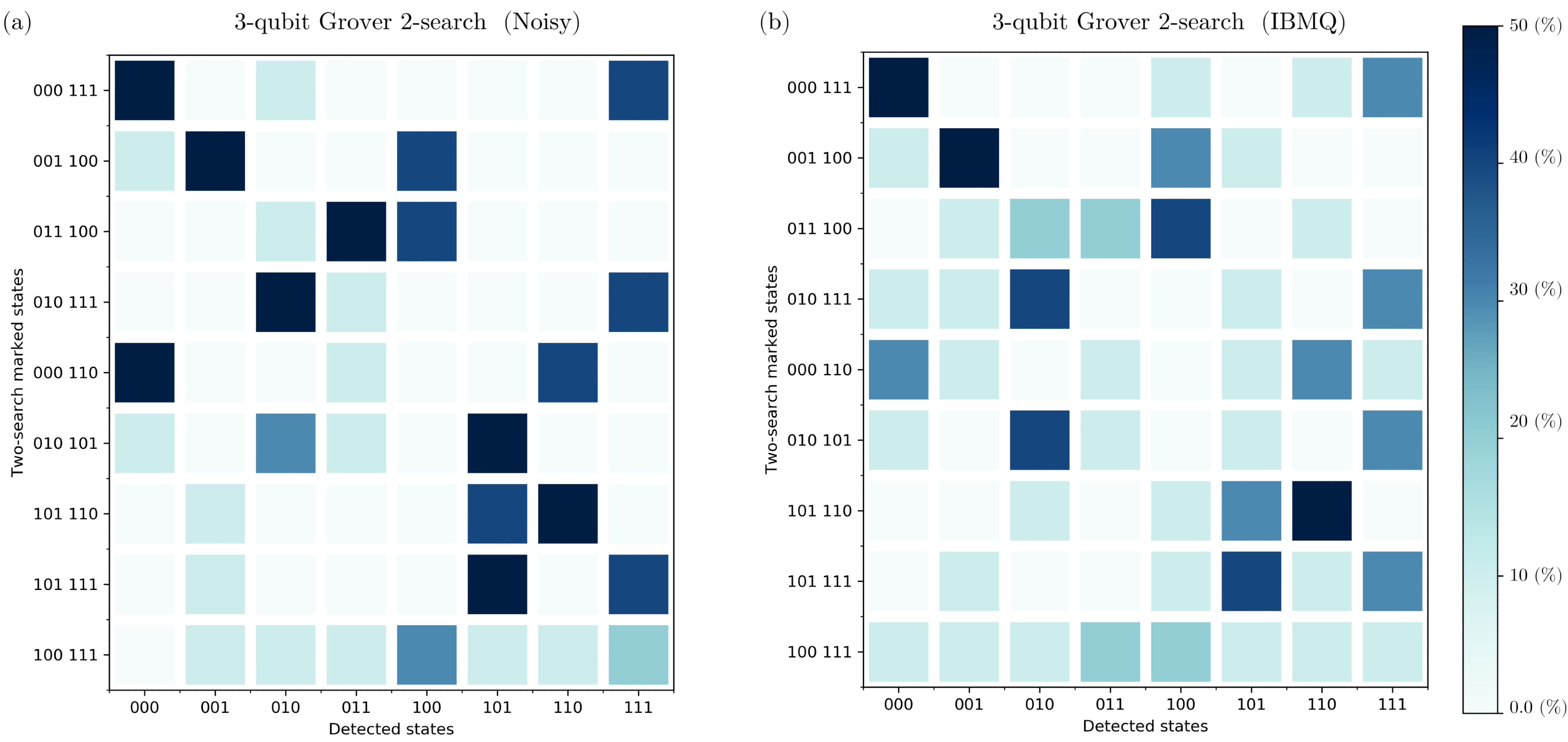}
\caption{Results derived from executing the GSA for nine two-solution scenarios in various environments on a 3-qubit database. The data from algorithm execution in a noisy environment is presented on the left, while data from execution on IBM Quantum's real quantum computers is displayed on the right. The graphs depict the probability distribution for each output state. We observed an average ${\cal ASP}$ of 84.44\% in the noisy execution and 64.44\% on the IBM quantum computers. Additionally, we obtained an average ${\cal SSO}$ of 84.03\% in the noisy environment and 63.10\%, on real IBM quantum computers. All percentages are calculated relative to the expected state $\ket{\psi_E}_\text{Multi}$, defined as $ \ket{\psi_E}_\text{Multi} = \frac{1}{\sqrt{2}} \ket{\chi^*_1} +\frac{1}{\sqrt{2}}  \ket{\chi^*_2}$, where  $\ket{\chi^*_1}$ and $\ket{\chi^*_2}$ represents the two marked states.
}
\label{fig:heat2}
\end{minipage}
\end{figure*}

For the single-solution scenarios, we observed intriguing outcomes (see Figure~\ref{fig:heat1}). In the presence of noise, our analysis revealed an average 
${\cal ASP}$ of \( 78.39\% \), indicating the algorithm's capability to consistently identify the correct solution amidst environmental disturbances. However, when executed on IBM Quantum's real quantum computers, the ${\cal ASP}$ decreased to \( 51.19\% \), underscoring the challenges encountered in practical quantum computing environments. 
Furthermore, our investigation into the 
${\cal SSO}$ metrics provided deeper insights. In noisy environments, we recorded an average ${\cal SSO}$ of \( 82.358\% \), reflecting the algorithm's ability to closely approximate the target state despite environmental noise. Conversely, on real IBM quantum computers, the average ${\cal SSO}$ decreased to \( 73.12\% \), indicating the degree of deviation from the expected state.

Transitioning to the two-solution scenarios, our findings revealed notable trends (see Figure~\ref{fig:heat2}). In noisy executions, the algorithm displayed a robust performance, achieving an average ${\cal ASP}$ of \( 84.44\% \). However, on IBM Quantum's real quantum computers, the ${\cal ASP}$ reduced to \( 64.44\% \), suggesting the impact of practical constraints on algorithmic efficacy. 
Examining the ${\cal SSO}$ values in these scenarios provided additional insights. In noisy environments, the average ${\cal SSO}$ was \( 84.03\% \), indicating a satisfactory overlap with the expected state despite noise. Conversely, on real IBM quantum computers, the average ${\cal SSO}$ decreased to \( 63.10\% \), highlighting the challenges encountered in achieving precise outcomes in practical quantum computing in the NISQ era.

\section{Experimental characterization of the Grover search algorithm} \label{Sec:3}

\subsection{${\cal QST}$ experiments}

${\cal QST}$ plays a pivotal role in the development and validation of quantum technologies by providing a comprehensive characterization of quantum states~\citep{EPJQ,QST_EPJ,Googles}. It allows extracting essential information about the state of a quantum system, 
enabling precise assessment of quantum operations' fidelity and performance~\citep{QST09,QST08,QST06,QST11,QST10,EPJQ,Googles}. By reconstructing quantum states experimentally, ${\cal QST}$ helps identify and quantify sources of errors, assess the effectiveness of error mitigation techniques, and verify the fidelity of quantum gates and algorithms~\citep{Entangling,QPT_fast,SQSCZ1,SQSCZ2}. Moreover, it serves as a crucial tool for benchmarking and calibrating quantum devices, facilitating progress towards achieving reliable and scalable quantum computation and communication protocols~\citep{Cprotocols,Qprotocols}.

Here, we undertake a comprehensive exploration through five ${\cal QST}$ experiments to meticulously evaluate the behavior and efficiency of our implemented algorithm under varied conditions. Specifically, we conduct two ${\cal QST}$ experiments for the GSA employing single search oracles—$\ket{010}$ and $\ket{101}$. Moreover, three additional ${\cal QST}$ experiments are meticulously performed for the GSA utilizing two search oracles—$\ket{000}-\ket{111}$, $\ket{101}-\ket{110}$, and $\ket{101}-\ket{111}$. These experiments are meticulously executed across three distinct environments: a pristine noise-free setting, a simulated noisy environment, and on IBM Quantum's tangible quantum computers.  For each setting, we measured the state fidelity of the output states produced by the algorithm. The state fidelity represents the similarity between the output states and the desired target states, $0 \leq {\cal F_S} \leq  1$, with higher fidelity values indicating better performance~\citep{F_State,Nilson}.

\begin{figure*}
\centering
\begin{minipage}{0.93\textwidth}
\centering
\includegraphics[width=0.49\textwidth]{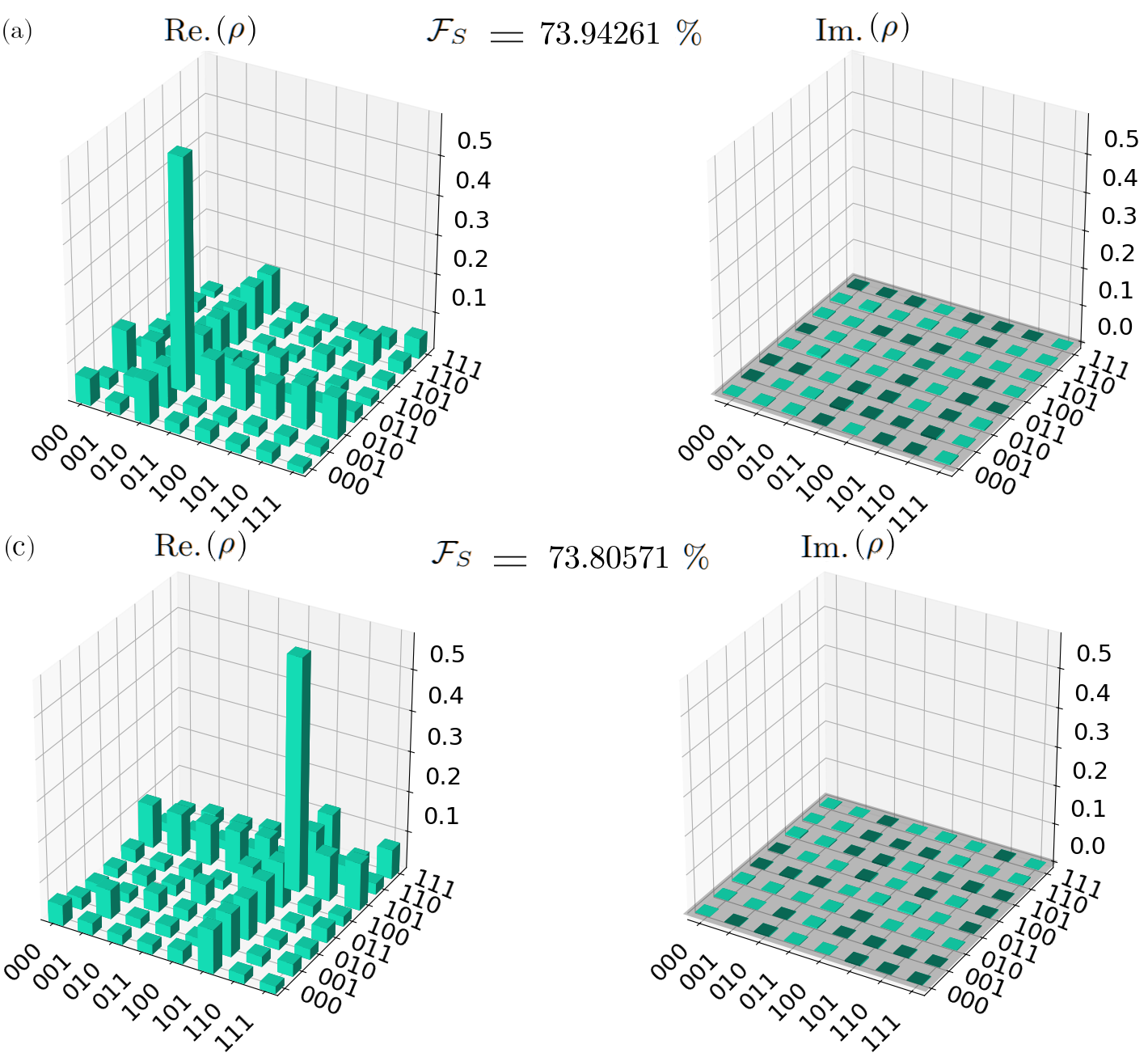}
\includegraphics[width=0.49\textwidth]{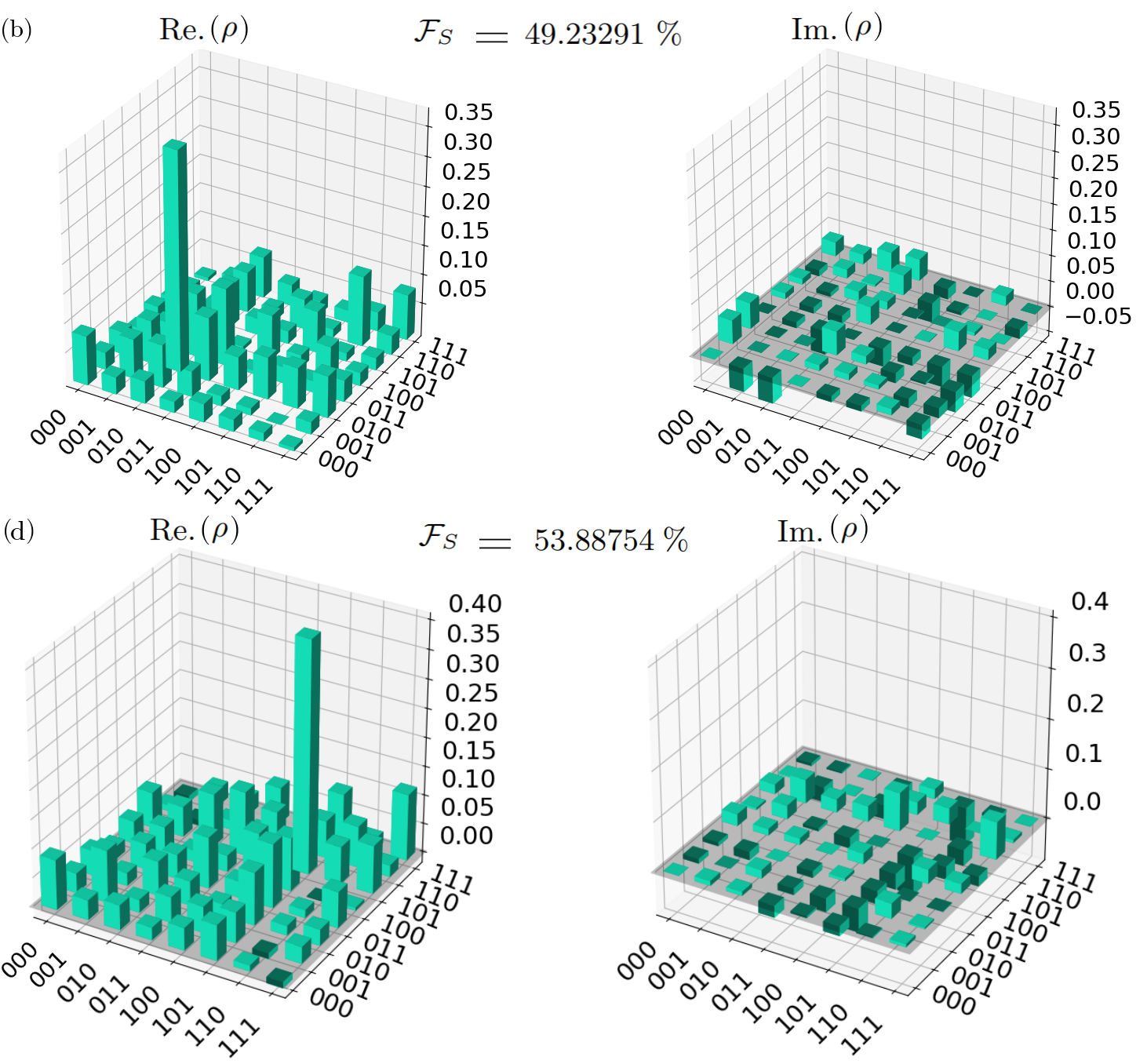}
\vspace{-0.5cm}
\caption{Results from ${\cal QST}$ experiments on the GSA with single-search oracles in different environments.
(a) Real (Re.$(\rho)$) and imaginary (Im.$(\rho)$) parts of the density matrix obtained from ${\cal QST}$ experiments for the single-search marked state $\ket{010}$, executed with 7797 repeated shots in a noisy environment, (b) and 1024 repeated shots on IBM Quantum's 127-qubit superconducting quantum computer \textit{ibm\_osaka}, yielding state fidelities (${\cal F}_S$) of 73.9426\% and 49.2329\%, respectively.
(c) Re.$(\rho)$ and Im.$(\rho)$ parts of the density matrix obtained from ${\cal QST}$ experiments for the single-search marked state $\ket{101}$, executed with 7797 repeated shots in a noisy environment, (d) and 7797 repeated shots on \textit{ibm\_osaka}, yielding ${\cal F}_S$ of 73.8057\% and 53.8875\%, respectively.}
\label{fig:QST1}
\end{minipage}
\begin{minipage}{0.93\textwidth}
\centering
\includegraphics[width=0.49\textwidth]{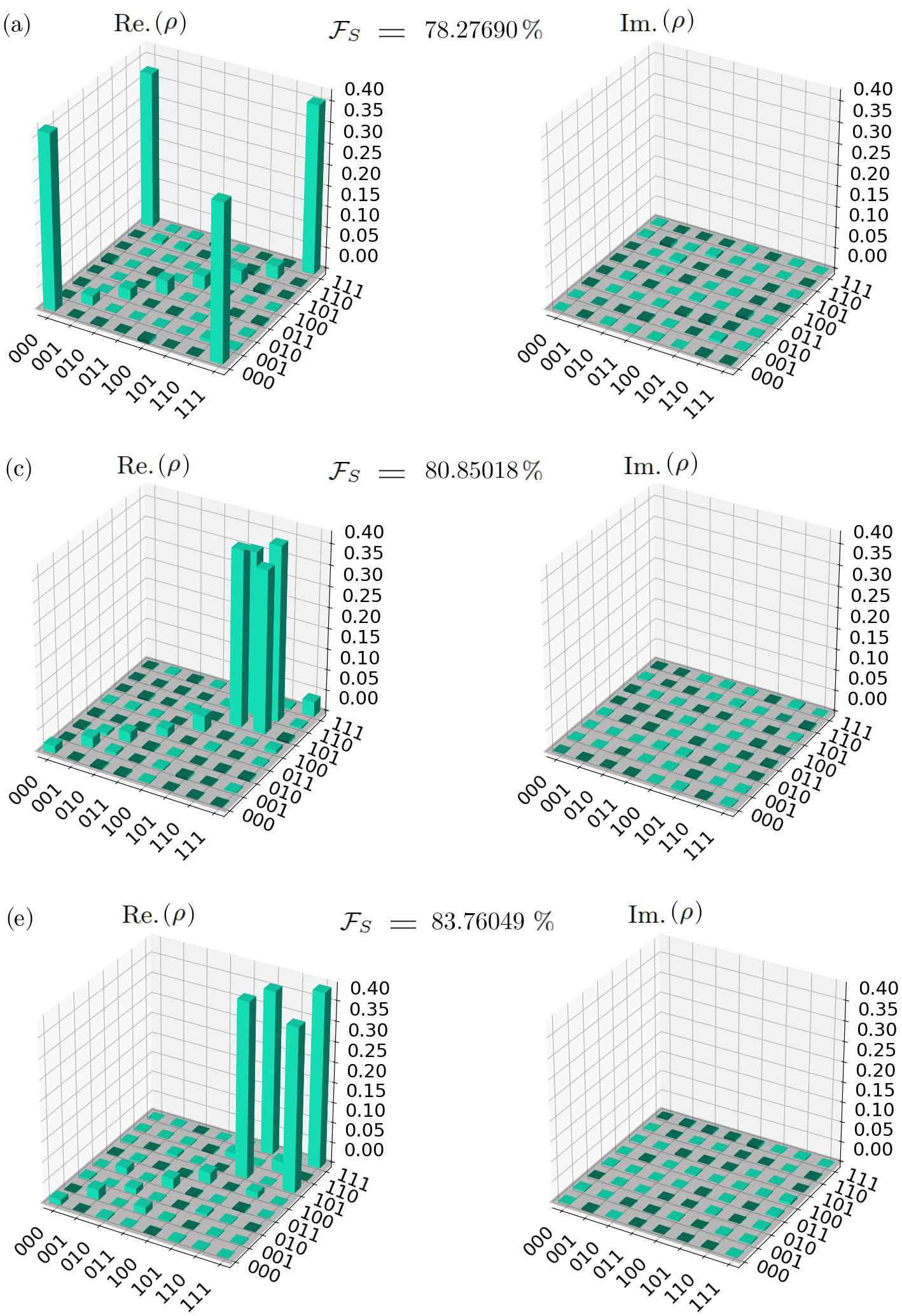} 
\includegraphics[width=0.49\textwidth]{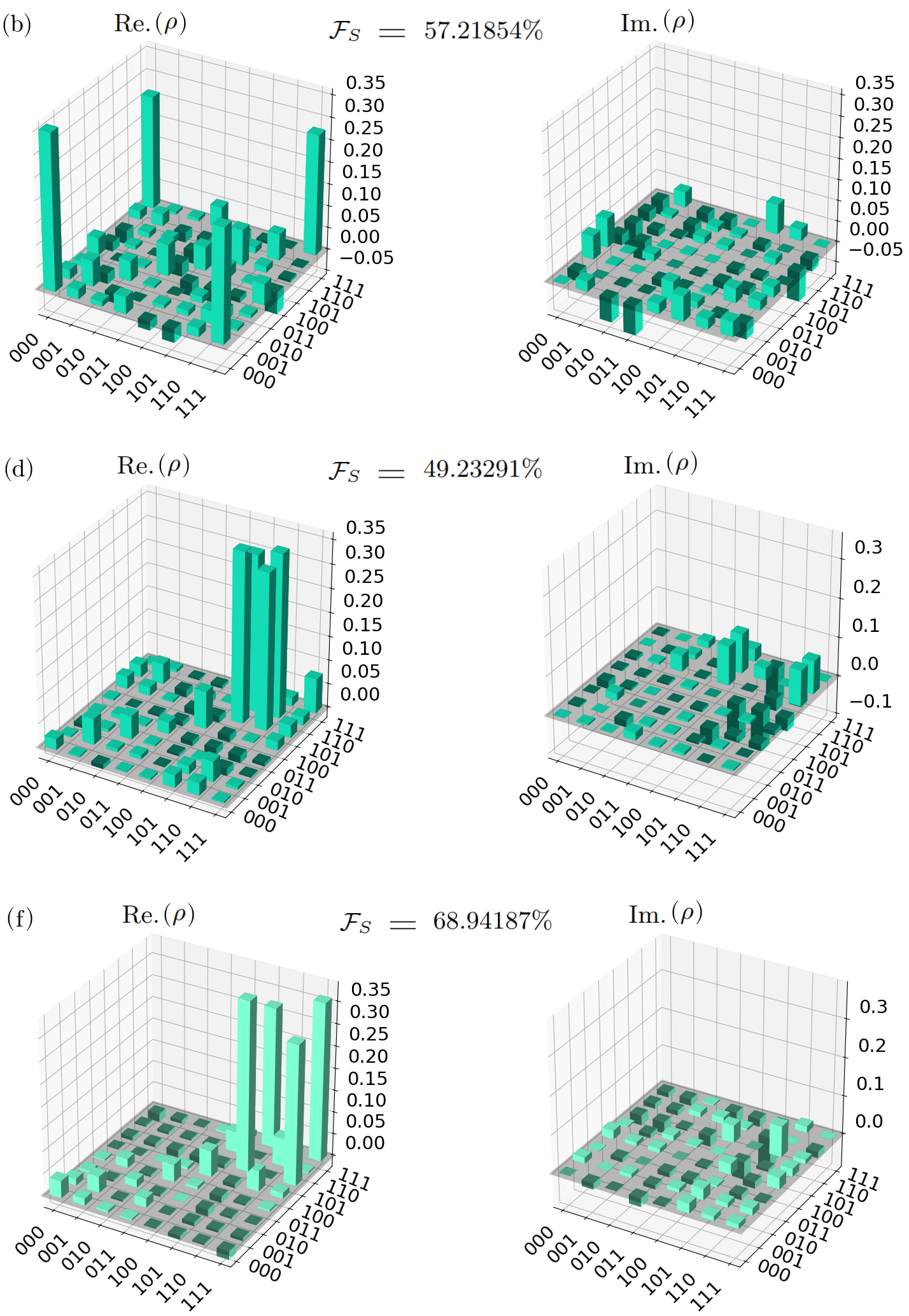} 
\vspace{-0.5cm}
\caption{Results from ${\cal QST}$ experiments of the GSA employing two-search oracles in different environments.
(a) Real (Re.$(\rho)$) and imaginary (Im.$(\rho)$) parts of the density matrix obtained from ${\cal QST}$ experiments for the two-search marked states $\ket{000}$ and $\ket{111}$, executed with 7797 repeated shots in a noisy environment, (b) and 7797 repeated shots on IBM Quantum's 127-qubit superconducting quantum computer \textit{ibm\_osaka}, yielding state fidelities (${\cal F}_S$) of 78.2769\% and 57.2185\%, respectively.
(c) Re.$(\rho)$ and Im.$(\rho)$ parts of the density matrix obtained from ${\cal QST}$ experiments for the two-search marked states $\ket{101}$ and $\ket{110}$, executed with 7797 repeated shots in a noisy environment, (d) and 1024 repeated shots on 
\textit{ibm\_osaka}, yielding 
${\cal F}_S$ of 80.8501\% and 49.2329\%, respectively.
(d) Re.$(\rho)$ and Im.$(\rho)$ parts of the density matrix obtained from ${\cal QST}$ experiments for the two-search marked states $\ket{111}$ and $\ket{101}$, executed with 7797 repeated shots in a noisy environment, (e) and 1024 repeated shots on \textit{ibm\_osaka}, yielding 
${\cal F}_S$ of 83.7605\% and 68.9419\%, respectively.
}
    \label{fig:QST2}
\end{minipage}
\end{figure*}

In the pursuit of authenticity, when executing our experiments on genuine quantum computers, we meticulously perform ${\cal QST}$ experiments for the GSA with single-marked states $\ket{010}$ and $\ket{101}$, employing 1024 and 7779 shots, respectively. 
The resultant state fidelities ${\cal F_S}$ are found to be 49.23291\% and 53.88754\%, respectively. Additionally, for ${\cal QST}$ experiments involving two-marked states (two search oracles)—$\ket{000}-\ket{111}$, $\ket{101}-\ket{110}$, and $\ket{101}-\ket{111}$—we replicate the experiments using 7779, 1024, and 1024 shots, respectively. The corresponding state fidelities are observed to be 57.21854\%, 49.23291\%, and 68.94187\%, respectively. 
The choice of varying numbers of shots for the ${\cal QST}$ experiments was employed to ensure sufficient statistical sampling and improve the reliability of the results. 
The outcomes gleaned from ${\cal QST}$ experiments of the 
GSA on real quantum computers, utilizing both single-search oracles and two-search oracles, are meticulously presented in Figure~\ref{fig:QST1} and Figure~\ref{fig:QST2}, respectively. 

Upon examining the results across the three distinct environments, a notable degradation in performance is observed as we transition from the noise-free setting (mean  ${\cal F_S} = 99.38 \%$) to the noisy setting (mean ${\cal F_S} = 78.13  \%$), and further to the real quantum computer setting (mean ${\cal F_S} = 54.32  \%$).  
In our ${\cal QST}$ experiments conducted on a real superconducting quantum computer for the GSA with different oracles, with a mean ${\cal F_S}$ of 
$54.32 \%$ and a standard deviation of 0.099, the algorithm demonstrates a moderate level of consistency in generating quantum states resembling the target state. These findings suggest that while the GSA shows promise in surpassing classical random search strategies, there remains room for improvement to achieve higher and more consistent fidelity. Further experimental investigation could elucidate potential avenues for refinement, ultimately advancing the algorithm's efficacy in quantum search applications.

\section{Analysis and discussions}\label{Sec:4}

In this section, following the collection of experimental data, we performed a comprehensive statistical analysis. Our aim was to derive meaningful conclusions regarding the performance of the GSA across various settings, focusing on aspects such as ${\cal ASP}$, ${\cal SSO}$, state fidelity ${\cal F}_S$, and to assess the influence of noise and other factors on its effectiveness.

\subsection{${\cal ASP}$ analysis}

The statistical tests conducted offer critical insights into the ${\cal ASP}$ of a 3-qubit GSA executed on IBM Quantum's real quantum computers, particularly in estimating the population of all 2-marked states oracles within the 3-qubit space (see Table~\ref{tab:single} and \ref{tab:multi}). The mean ${\cal ASP}$ of 64.44\% provides a central measure of the algorithm's performance in identifying desired states within the quantum system, while the standard deviation of 16.67 offers an indication of the variability in ${\cal ASP}$ values around the mean. The subsequent one-sample t-test comparing the observed mean ${\cal ASP}$ against a hypothesized population mean yields a non-significant p-value of 0.99, indicating that the observed mean ${\cal ASP}$ is likely representative of the population mean. This suggests that the algorithm's performance, as observed on IBM Quantum's real quantum computers, aligns closely with theoretical expectations or benchmark values. Similarly, the hypothesis test for variance, with a p-value of 0.868, suggests no significant difference in the variability of ${\cal ASP}$ between the observed sample and the hypothetical population. Thus, the variability in ${\cal ASP}$ observed in the sample is deemed consistent with that of the population. Overall, these results underscore the reliability and validity of the observed ${\cal ASP}$ values in estimating the population of 2-marked states within the 3-qubit space.

\subsection{${\cal SSO}$ analysis}

The statistical tests conducted on the ${\cal SSO}$ for a 3-qubit GSA executed on IBM Quantum's real quantum computers provide reliable insights into estimating population parameters for all 2-marked states oracles within the 3-qubit space (see Table~\ref{tab:single} and \ref{tab:multi}). The mean ${\cal SSO}$ value of 63.10 represents the average squared statistical overlap between observed two-search result oracles, with a standard deviation of 17.07 indicating variability around this mean. A one-sample t-test with a 95\% confidence interval (CI), yielding a non-significant p-value of 0.99, suggests that the observed mean ${\cal SSO}$ is representative of the population mean. Additionally, the 95\% confidence intervals for variance offer insights into the precision of estimated variability within the population. The hypothesis test for variance, with a p-value of 0.867, indicates no significant difference in variability between the observed sample and the hypothetical population. These results underscore the accuracy and reliability of the observed ${\cal SSO}$ values in estimating population parameters.

\begin{table*}
    \centering
    \caption{\label{tab:single}Comparative analysis of ${\cal SSO}$ and ${\cal ASP}$ for all 8 ($2^n$) single-search result oracles in the GSA. The experiments are conducted under diverse conditions: noise-free, noisy, and on IBM Quantum's 127-qubit superconducting quantum computers. 
    \footnote{Data from the noise-less environment is not presented here, as it closely approaches 99.99\%. For a comparison of the GSA performance across various environments, see Appendix~\ref{A:1}.}}
    \begin{tabular}{l|c|c|c|c}
    \hline \hline
        Marked State &  ${\cal SSO}$ (Noisy setting)  & ${\cal SSO}$ (IBM Quantum)  & ${\cal ASP}$ (Noisy setting)  & ${\cal ASP}$ (IBM Quantum)  \\
        \hline \hline
        (1) 000 &82.49  (\%) &74.25 (\%) &76.79 (\%) &51.20 (\%) \\
        (2) 001 &82.30  (\%) &82.30 (\%) &89.59 (\%) &89.59 (\%) \\
        (3) 010 &81.81  (\%) &71.01 (\%) &64.00 (\%) &38.39 (\%) \\
        (4) 011 &82.30  (\%) &64.73 (\%) &89.59 (\%) &38.39 (\%) \\
        (5) 100 &82.49  (\%) &74.25 (\%) &76.79 (\%) &51.20 (\%) \\
        (6) 101 &82.49  (\%) &81.81 (\%) &76.79 (\%) &64.00 (\%) \\
        (7) 110 &82.49  (\%) &65.60 (\%) &76.79 (\%) &38.39 (\%) \\
        (8) 111 &82.49  (\%) &71.01 (\%) &76.79 (\%) &38.39 (\%) \\
        Median       &82.49 (\%)  &72.63  (\%) &76.79 (\%) &44.80 (\%) \\ 
        StDev        &0.237       &6.53        &8.20       &18.10   \\
        Mean         &82.358 (\%) &73.12  (\%) &78.39 (\%) &51.19 (\%) \\
        SE Mean\footnotemark[2]      &0.0839      &2.31        &2.90       &6.40  \\
        \hline \hline
    \end{tabular}
    \footnotetext[2]{The standard error of the mean ($\text{SE Mean}=\text{StDev}/\sqrt{S}$).}
\end{table*}

\begin{table*}
    \centering
   \caption{\label{tab:multi}Comparative Analysis of ${\cal SSO}$ and ${\cal ASP}$ for nine two-search result oracles in the GSA. The experiments are conducted under various conditions: noise-free, noisy, and on IBM Quantum's 127-qubit superconducting quantum computers. Additionally, the analysis incorporates hypothesis testing and $95\%$ CI for both the population mean ($\mu$) and variance ($\sigma^2$).}
    \begin{tabular}{l|c|c|c|c}
    \hline \hline
        Two-Marked State &  ${\cal SSO}$ (Noisy setting)  & ${\cal SSO}$ (IBM Quantum)   &  ${\cal ASP}$ (Noisy setting)   & ${\cal ASP}$ (IBM Quantum) \\
        \hline \hline
        (1) 000, 111 &89.72  (\%) &78.72 (\%) &90.00 (\%) &80.00 (\%) \\
        (2) 001, 100 &89.72  (\%) &78.72 (\%) &90.00 (\%) &80.00 (\%) \\
        (3) 011, 100 &89.72  (\%) &58.28 (\%) &90.00 (\%) &60.00 (\%) \\
        (4) 010, 101 &89.72  (\%) &69.64 (\%) &90.00 (\%) &70.00 (\%) \\
        (5) 000, 110 &89.72  (\%) &60.00 (\%) &90.00 (\%) &60.00 (\%) \\
        (6) 010, 101 &78.73  (\%) &45.00 (\%) &80.00 (\%) &50.00 (\%) \\
        (7) 101, 110 &89.72  (\%) &78.72 (\%) &90.00 (\%) &80.00 (\%) \\
        (8) 101, 111 &89.72  (\%) &69.64 (\%) &90.00 (\%) &70.00 (\%) \\
        (9) 100, 111 &49.49  (\%) &29.14 (\%) &50.00 (\%) &30.00 (\%)\\
        Mean              &84.03 (\%)  &63.10 (\%)  &84.44 (\%)  &64.44 (\%)\\
        StDev             &13.45       &17.07       &13.33      &16.67   \\
        SE Mean           &4.48        &5.69        &4.44       &5.56  \\
        t-test (95\% CI)  &(73.69,\, 94.37)      &(49.97,\,76.22)   &(74.20,\,94.69)  &(51.63,\,77.26)  \\
        Hypothesis tests (Mean\footnotemark[2])  &P-value: 1.00     & 0.99  & 0.99  & 0.99  \\
        95\% CI StDev     &(9.10,\, 25.8)         &(11.5,\,32.7)    &(9.00,\,25.5)  &(11.3,\,31.9)\\
        95\% CI Variance  &(83,\, 664)           &(133,\,1070)      &(81,\,625)  &(127,\,1019) \\
        Hypothesis tests (Variance\footnotemark[3])  &P-value: 0.866  & 0.867  & 0.866   & 0.868  \\
        \hline \hline
    \end{tabular}
\footnote[2]{The null and alternative hypothesis for the mean are: \\
$\mu=84.03$ vs $\mu \neq 84.03$, \\
$\mu=63.10$ vs $\mu \neq 63.10$, \\
$\mu=84.44$ vs $\mu \neq 84.44$, and \\
$\mu=64.44$ vs $\mu \neq 64.44$, respectively.}
\footnote[3]{The null and alternative hypothesis for the variance are: \\
$\sigma=13.45$ vs $\sigma \neq 13.45$,\\ 
$\sigma=17.07$ vs $\sigma \neq 17.07$, \\
$\sigma=13.33$ vs $\sigma \neq 13.33$, and \\
$\sigma=16.67$ vs $\sigma \neq 16.67$, respectively.}    
\end{table*}

\begin{table*}
    \centering
    \caption{\label{tab:qst}Analysis of state fidelity (${\cal F}_S$) derived from ${\cal QST}$ experiments conducted on the GSA, encompassing both single-search and two-search oracles. The experiments are performed across various environments: Noise-free, Noisy, and on IBM Quantum's 127-qubit superconducting quantum computers. Furthermore, the analysis includes hypothesis testing and $95\%$ CI for the population mean ($\mu$) and variance ($\sigma^2$).}
    \begin{tabular}{l|c|c|c}
        \hline \hline
        Marked State & ${\cal F}_S$ ${\cal QST}$ (Noise-free)& ${\cal F}_S$ ${\cal QST}$ (Noisy settings) & ${\cal F}_S$ ${\cal QST}$ (IBM Quantum) \\
        \hline \hline
        Single-search (010)         &0.9946673  &0.7394261   &0.4923291  \\
        Single-search (101)         &0.9946291  &0.7380571   &0.5388754  \\
        Two-search (000, 111)    &0.9956002  &0.7827690   &0.5721854  \\
        Two-search (101, 110)    &0.9922922  &0.8085018   &0.4923291  \\
        Two-search (111, 101)    &0.9921588  &0.8376049   &0.6894187  \\
        Mean         &0.993870   &0.7813   &0.5432\\
        StDev        &0.001551   &0.0434   &0.0990\\
        SE Mean      &0.000694   &0.01194  &0.0443\\
        t-test (95\% CI)  &(0.9919,\, 0.9957)     &(0.7274,\, 0.8352)    &(0.4205,\, 0.6661)    \\
        Hypothesis tests (Mean\footnotemark[1])  & P-value: 0.999 &  0.999 &1.000 \\
        95\% CI StDev     &(0.00093,\, 0.00446)   &(0.0260,\,  0.1247)  &(0.05930,\, 0.02845)      \\
        95\% CI Variance  &(0.000001,\, 0.00002)   &(0.00068,\, 0.01556)  &(0.00352,\, 0.08093)       \\
        Hypothesis tests (Variance\footnotemark[2])  &P-value: 0.812 & 0.812 & 0.812 \\
        \hline \hline
    \end{tabular}
\footnote[1]{The null and alternative hypothesis for the mean are: \\
$\mu=0.9938$ vs $\mu \neq 0.9938$, \\
$\mu=0.7813$ vs $\mu \neq 0.7813$, and \\
$\mu=0.5432$ vs $\mu \neq 0.5432$, respectively.}
\footnote[2]{The null and alternative hypothesis for the variance are: \\
$\sigma=0.00155$ vs $\sigma \neq 0.00155$, \\
$\sigma=0.04340$ vs $\sigma \neq 0.0434$, and \\
$\sigma=0.09900$ vs $\sigma \neq 0.099$, respectively.}    
\end{table*}

\subsection{Evaluation of state fidelity}

We conducted a statistical analysis to assessing the reliability of the state fidelity measurement 
and providing insights into the variability of the fidelity of the GSA with different marked states and to examine the differences in algorithm performance across the three settings (as presented in Table~\ref{tab:qst}), aiming to understand the fidelity of the population of all 3-qubit marked states.

The statistical tests on the state fidelity of a 3-qubit GSA executed on real quantum computers provide valuable insights into the fidelity of all 3-qubit marked states. The mean fidelity of 0.5432, with a standard deviation of 0.099, reflects both the average fidelity and its variability across experiments. 
The one-sample t-test with a 95\% CI (0.4205, 0.6661) compares the observed mean fidelity against a hypothesized population mean. The resulting p-value of 1.00 indicates that there is insufficient evidence to reject the null hypothesis, implying that the observed mean fidelity is not significantly different from the hypothesized population mean. This suggests that the observed mean fidelity is likely representative of the population. 
Similarly, the hypothesis test for variance, yielding a p-value of 0.812, suggests no significant difference in variability between the observed sample and the hypothetical population.

This reinforces the reliability of the observed sample's representation of the population and underscores the consistency in fidelity across different experiments. 
These tests provide confidence in the accuracy and reliability of the observed ${\cal F}_S$ values, aiding in understanding the fidelity of the population of all 3-qubit marked states.

\section{Conclusion}\label{Sec:5}

The Grover search algorithm stands as a pivotal advancement in the field of quantum computing, offering a revolutionary approach to solving unstructured search problems. This algorithm addresses a fundamental challenge in computing—finding a desired item in an unsorted database significantly faster than classical methods allow. Its importance cannot be overstated, as it promises exponential speedup over classical algorithms for a wide range of applications, from cryptography to database search and optimization.

Our comprehensive study sheds light on the practical implementation and performance evaluation of GSA on real 127-qubits superconducting quantum computers. Through our exploration of single-solution and two-solution scenarios, we have uncovered both promising capabilities and significant challenges. Despite the presence of noise, the algorithm demonstrated a remarkable ability to maintain a high ${\cal ASP}$ and ${\cal SSO}$. For instance, in single-solution scenarios, we observed an average ${\cal ASP}$ of 78.39\% and an average ${\cal SSO}$ of 82.358\%, and in two-solution scenarios, an average ${\cal ASP}$ of 84.44\% and an average ${\cal SSO}$ of 84.03\%. 
However, executing the algorithm on real superconducting quantum computers revealed practical constraints, leading to decreased ${\cal ASP}$ and ${\cal SSO}$ metrics. Specifically, we noted an average ${\cal ASP}$ of 51.19\% and an average ${\cal SSO}$ of 73.12\% in single-solution scenarios, and an average ${\cal ASP}$ of 64.44\% and an average ${\cal SSO}$ of 63.10\% in two-solution scenarios.

Furthermore, our statistical analyses, which included one-sample t-tests and 95\% confidence intervals, provided robust insights into the consistency and reliability of the performance metrics we observed. The one-sample t-tests comparing the observed mean ${\cal ASP}$ and ${\cal SSO}$ against hypothesized population means yielded non-significant p-values, indicating that the observed means of ${\cal ASP}$ and ${\cal SSO}$ are statistically representative of the population means in both scenarios. Similarly, hypothesis tests for variance produced p-values of 0.868 and 0.867, suggesting that there is no significant difference in the variability of ${\cal ASP}$ and ${\cal SSO}$ between our observed sample and the hypothetical population.

Additionally, we conducted five ${\cal QST}$ experiments on IBM Quantum's 127-qubits superconducting quantum computer, \textit{ibm\_osaka}, to assess the fidelity of output states in both single-search and two-search scenarios of the GSA. Our experiments using ${\cal QST}$ provided strong confidence in the fidelity of output states. Statistical tests on the state fidelity (${\cal F}_S$) of a 3-qubit GSA executed on real quantum computers revealed a mean ${\cal F_S}$ of 0.5432. A one-sample t-test with a 95\% confidence interval (0.4205, 0.6661) comparing the observed mean state fidelity against a hypothesized population mean yielded a non-significant p-value, indicating the observed state fidelity is likely representative of the population, with no significant variance observed, suggesting no significant difference in variability between the observed sample and the hypothetical population, reinforcing the reliability and consistency of fidelity across experiments.

By addressing practical challenges such as noise and environmental disturbances, we can further enhance the scalability and applicability of quantum algorithms in real-world settings. Our research endeavored to contribute to the ongoing efforts to bridge the gap between theoretical developments and practical implementations, paving the way for transformative applications across various domains.

\appendix

\section{Quantum Hardware}\label{A:1}

\subsection{Algorithm Performance}

\begin{figure*}
    \centering
       \includegraphics[width=0.9\textwidth]{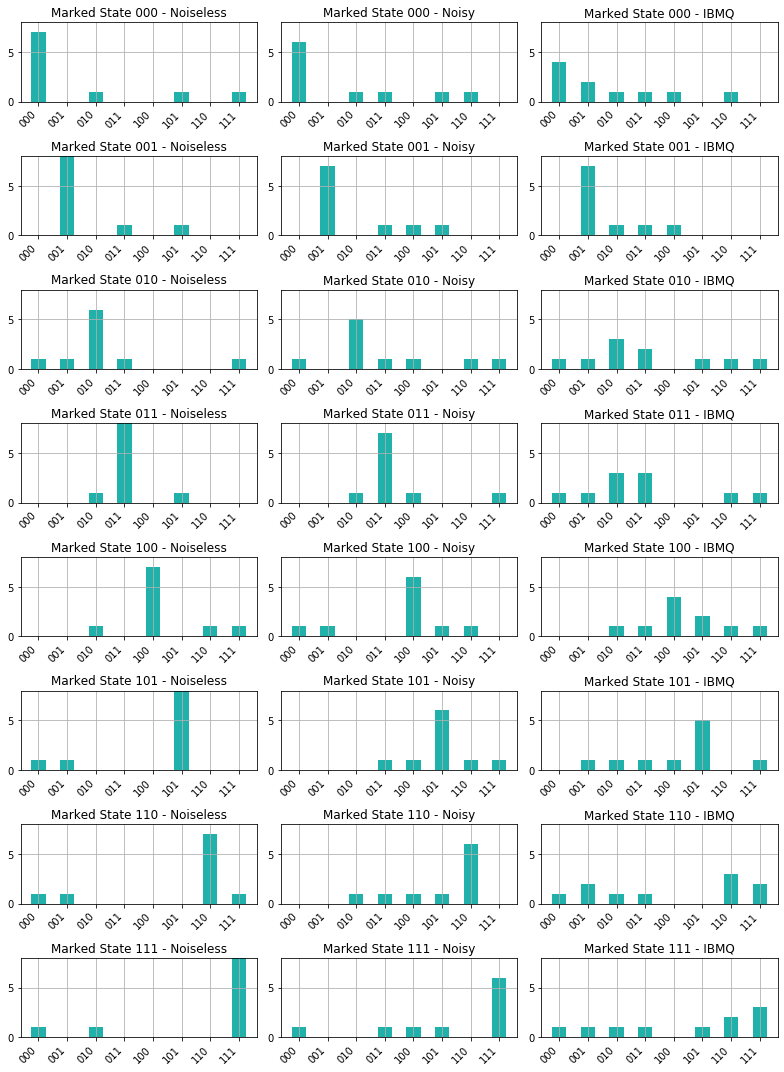}
    \caption{
    GSA performance across all 3-qubit single-marked states 
($\ket{000}, \, \ket{001}$, 
$\ket{010}, \, \ket{011}$, 
$\ket{100}, \, \ket{101}$, 
$\ket{110}, \, \text{and} \ket{111}$) across different environments: noise-free, noisy environment, and IBM Quantum's quantum hardware.}
    \label{fig:perf1}
\end{figure*}

\begin{figure*}
    \centering
       \includegraphics[width=0.9\textwidth, height=\textheight]{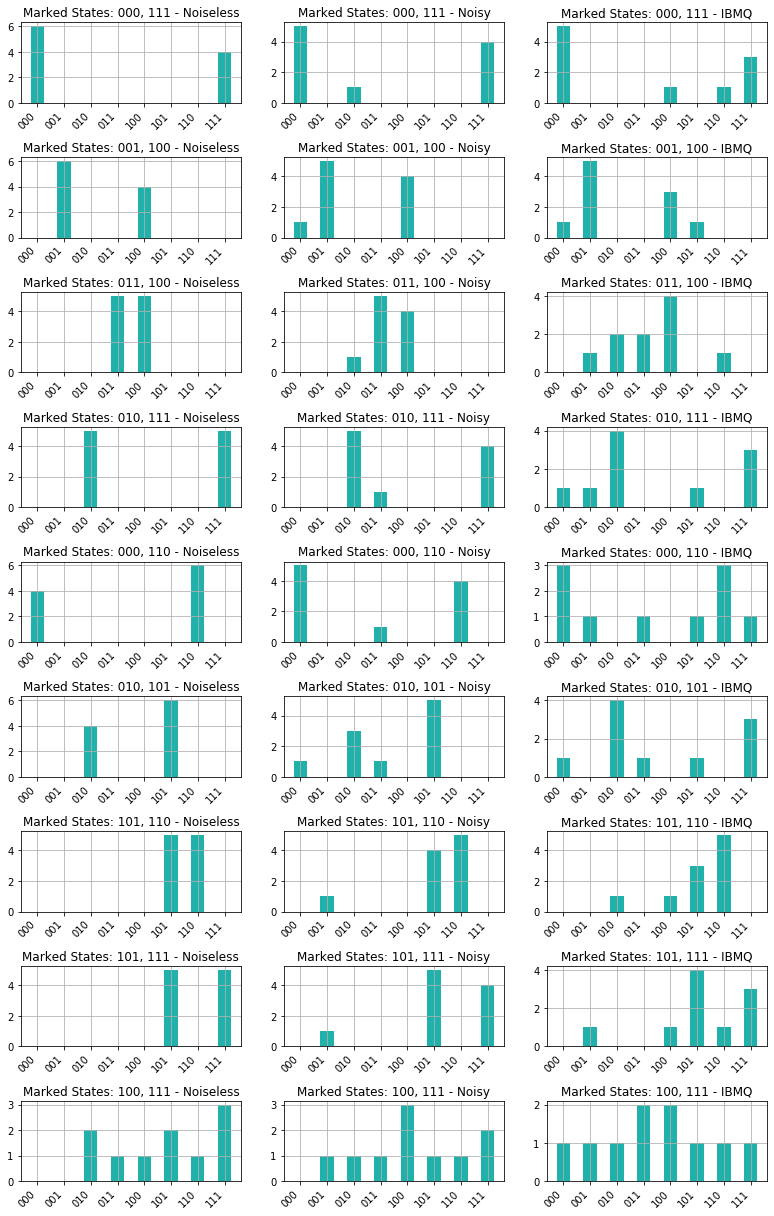}
    \caption{ 
    GSA performance for the nine 3-qubit two-marked states 
($\ket{000}-\, \ket{111}$,
$\ket{001}-\, \ket{100}$,
$\ket{011}-\, \ket{100}$,
$\ket{010}-\, \ket{111}$,
$\ket{000}-\, \ket{110}$,
$\ket{010}-\, \ket{101}$,
$\ket{101}-\, \ket{110}$,
$\ket{101}-\, \ket{111}$, and
$\ket{100}-\, \ket{111}$) across different environments: noise-free, noisy environment, and IBM Quantum's quantum hardware.}
    \label{fig:perf2}
\end{figure*}

The performance of the GSA across different environments is depicted in Figure~\ref{fig:perf1} and Figure~\ref{fig:perf2}. 
Figure~\ref{fig:perf1} illustrates the performance of the GSA across all 3-qubit single-marked states ($\ket{000}, \, \ket{001}$, 
$\ket{010}, \, \ket{011}$, 
$\ket{100}, \, \ket{101}$, 
$\ket{110}, \, \text{and} \ket{111}$) across different environments: noise-free, noisy Simulation, and IBM Quantum’s quantum hardware. 
While Figure~\ref{fig:perf2} showcases the performance for the nine 3-qubit two-marked states (
$\ket{000}-\, \ket{111}$,
$\ket{001}-\, \ket{100}$,
$\ket{011}-\, \ket{100}$,
$\ket{010}-\, \ket{111}$,
$\ket{000}-\, \ket{110}$,
$\ket{010}-\, \ket{101}$,
$\ket{101}-\, \ket{110}$,
$\ket{101}-\, \ket{111}$, and
$\ket{100}-\, \ket{111}$), evaluating their performance across the noise-free, noisy, and IBM Quantum’s quantum hardware environments.

\subsection{Hardware Specifications}\label{A1:2}

In this section, we provide a summary of key specifications and qubit characteristics for three state-of-the-art 127-qubit superconducting quantum computers utilized in our experimental endeavors within this research. These cutting-edge quantum machines, developed by IBM Quantum, namely: \textit{ibm\_sherbrook}, \textit{ibm\_osaka}, and \textit{ibm\_kyoto}. 
Table~\ref{tab:sherbrook} presents the specifications of \textit{ibm\_sherbrook}, including its model, basis gates, processor type, median ECR error, and CLOPS. The processor type is Eagle r3, and the quantum system exhibits a median ECR error of $7.565 \times 10^{-3}$, with 5000 CLOPS. 
While Table~\ref{tab:osaka} and~\ref{tab:kyoto} provide a detailed summary of the essential specifications and qubit characteristics for the 127-qubit quantum computers, \textit{ibm\_osaka} and \textit{ibm\_kyoto}, respectively.

\begin{table*}
\begin{minipage}{\textwidth}
\caption{ \label{tab:sherbrook} 
 Summary of key specifications and qubit characteristics for the IBM Quantum computer with 127 qubits, named \textit{ibm\_sherbrook}. Error rates for individual gates and readout are provided, with basis gates including ECR, ID, RZ, SX, and X. The processor type is Eagle r3 (version 1.4.49). This quantum system exhibits median errors: ECR error of $7.565\times 10^{-3}$, SX error of $2.093 \times 10^{-4}$, and readout error of $1.370 \times 10^{-2}$. The median $T_1$ is 264.82 $\mu$s, and the median $T_2$ is 185.58 $\mu$s, accompanied by a performance metric of 5,000 CLOPS (CLOPS: circuit layer operations per second). 
 Most systems support a maximum of 300 circuits and 100,000 shots, with some limitations, allowing a maximum of 100 circuits and 20,000 shots. Accessed May 18, 2024.}
\centering 
\begin{ruledtabular}
    \begin{tabular}{l| c| c |c |c| c| c |c |c |c |c |c}
      Properties& \( T_1 \) ($\mu$s) & \( T_2 \) ($\mu$s) & \( f_{\text{qubit}} \)(GHz) & \( \Delta \) (GHz) &  \( \epsilon_{\text{readout}} \) & \( P_{\text{m0p1}} \) & \( P_{\text{m1p0}} \) & \( t_{\text{readout}} \) (ns) & \( \epsilon_{\text{ID}} \) & \( \epsilon_{\text{SX}} \) & \( \epsilon_{\text{Pauli-X}} \) \\
      \hline   \hline
Mean                &263.9271  &183.6175   &4.7899  &-0.31039  &0.02925  &0.0354 &0.0230  &1244.44  &0.00046  &0.00046 &0.00046\\  
StDev               & 91.0896  &106.6288   &0.1091  & 0.00548  &0.05786  &0.0745 &0.0508  &0.0      &0.00094  &0.00094 &0.00094\\  
Min.                &  6.7271  &  6.7600   &4.4552  &-0.32426  &0.00280  &0.0034 &0.0014  &1244.44  &0.00010  &0.00010 &0.00010\\  
25\%                &203.1046  & 99.3026   &4.7315  &-0.31225  &0.00720  &0.0078 &0.0051  &1244.44  &0.00016  &0.00016 &0.00016\\  
Median              &255.9965  &176.4739   &4.7940  &-0.31093  &0.01130  &0.0128 &0.0090  &1244.44  &0.00022  &0.00022 &0.00022\\  
75\%                &336.8686  &237.9368   &4.8593  &-0.30962  &0.02465  &0.0296 &0.0180  &1244.44  &0.00032  &0.00032 &0.00032\\  
Max.                &466.3459  &571.7688   &5.0575  &-0.27186  &0.40130  &0.4833 &0.3924  &1244.44  &0.00618  &0.00618 &0.00618\\
    \end{tabular}%
\footnote{Definitions and characterization of key quantum parameters in our experiments: 
\(T_1\)- the relaxation time of a qubit, measured in microseconds (\(\mu s\)). 
\(T_2\)- the coherence time of a qubit, measured in microseconds (\(\mu s\)). 
\( f_{\text{qubit}} \)- qubit frequency in (GHz). 
\( \Delta \)- qubit anharmonicity in (GHz). 
\( \epsilon_{\text{readout}} \)- readout assignment error. 
\( P_{\text{m0p1}} \)-  prob meas0 prep1. 
\( P_{\text{m1p0}} \)- prob meas1 prep0. 
\( t_{\text{readout}} \)- readout length measured in (ns). 
\( \epsilon_{\text{ID}} \)- ID error. 
 \( \epsilon_{\text{SX}} \)- $\sqrt{X}$ or (SX) error. 
 \( \epsilon_{\text{Pauli-X}} \)- Pauli-X error.
}
\end{ruledtabular}
\vspace{0.7em}
 \caption{ \label{tab:osaka} Summary of key specifications and qubit characteristics for the IBM Quantum computer with 127 qubits, named \textit{ibm\_osaka}. Error rates for individual gates and readout are provided, with basis gates including ECR, ID, RZ, SX, and X. The processor type is Eagle r3 (version 1.1.7). This quantum system exhibits median errors: ECR error of $9.291 \times 10^{-3}$, SX error of $2.972 \times 10^{-4}$, and readout error of $2.320 \times 10^{-2}$. The median $T_1$ is 265.09 $\mu$s, and the median $T_2$ is 118.88 $\mu$s, accompanied by a performance metric of 5,000 CLOPS. Accessed May 27, 2024.}
\centering 
\begin{ruledtabular}
\begin{tabular}{l| c| c |c |c| c| c |c |c |c |c |c}
Properties& \( T_1 \) ($\mu$s) & \( T_2 \) ($\mu$s) & \( f_{\text{qubit}} \)(GHz) & \( \Delta \) (GHz) &  \( \epsilon_{\text{readout}} \) & \( P_{\text{m0p1}} \) & \( P_{\text{m1p0}} \) & \( t_{\text{readout}} \) (ns) & \( \epsilon_{\text{ID}} \) & \( \epsilon_{\text{SX}} \) & \( \epsilon_{\text{Pauli-X}} \) \\
      \hline   \hline      
Mean            &272.0908   &156.9850  &4.8542   &-0.28497  &0.0523  &0.0529   &0.0517  &1400.00  &0.00228  &0.00228  &0.00228 \\
StDev           &98.88077   &100.2290  &0.1144   &0.079198  &0.0784  &0.0837   &0.0857  &  0.0   &0.01049  &0.01049  &0.01049 \\
Min.            &7.717160   &5.952153  &4.5680   &-0.31199  &0.0035  &0.0040   &0.0014  &1400.00  &0.00009  &0.00009  &0.00009 \\
25\%            &200.5397   &76.38629  &4.7723   &-0.30888  &0.0110  &0.0117   &0.0101  &1400.00  &0.00016  &0.00016  &0.00016 \\
\text{Median}   &280.8633   &147.2738  &4.8612   &-0.30749  &0.0231  &0.0220   &0.0208  &1400.00  &0.00025  &0.00025  &0.00025 \\
75\%            &342.0401   &242.5020  &4.9283   &-0.30611  &0.0591  &0.0588   &0.0614  &1400.00  &0.00043  &0.00043  &0.00043 \\
Max.            &469.1791   &384.6830  &5.1283   &0.0       &0.4931  &0.5000   &0.6896  &1400.00  &0.08711  &0.08711  &0.08711 \\ 
    \end{tabular}%
\end{ruledtabular}
\vspace{0.7em}
\caption{\label{tab:kyoto} Summary of key specifications and qubit characteristics for the IBM Quantum computer with 127 qubits, named \textit{ibm\_kyoto}. Error rates for individual gates and readout are provided, with basis gates including ECR, ID, RZ, SX, and X. The processor type is Eagle r3 (version 1.2.38). This quantum system exhibits median errors: ECR error of $9.675 \times 10^{-3}$, SX error of $3.080 \times 10^{-4}$, and readout error of $1.660 \times 10^{-2}$. The median $T_1$ is 215.71 $\mu$s, and the median $T_2$ is 90.64 $\mu$s, accompanied by a performance metric of 5,000 CLOPS. Accessed May 27, 2024.}
\centering 
\begin{ruledtabular}
\begin{tabular}{l| c| c |c |c| c| c |c |c |c |c |c}
Properties& \( T_1 \) ($\mu$s) & \( T_2 \) ($\mu$s) & \( f_{\text{qubit}} \)(GHz) & \( \Delta \) (GHz) &  \( \epsilon_{\text{readout}} \) & \( P_{\text{m0p1}} \) & \( P_{\text{m1p0}} \) & \( t_{\text{readout}} \) (ns) & \( \epsilon_{\text{ID}} \) & \( \epsilon_{\text{SX}} \) & \( \epsilon_{\text{Pauli-X}} \) \\
      \hline   \hline      
Mean              &215.8543  &117.8829 &4.9666 &-0.2928  &0.0361  &0.0375  &0.0346  &1400.00  &0.00306  &0.00306 &0.00306  \\
StDev             & 71.4440  & 85.6455 &0.1301 & 0.0654  &0.0508  &0.0633  &0.0462  &   0.0  &0.02250  &0.02250 &0.02250  \\
Min.              &  0.8683  &  3.6123 &4.7045 &-0.3120  &0.0026  &0.0030  &0.0022  &1400.00  &0.00009  &0.00009 &0.00009  \\   
25\%              &173.8905  & 47.1228 &4.8574 &-0.3087  &0.0096  &0.0092  &0.0083  &1400.00  &0.00019  &0.00019 &0.00019  \\ 
Median            &221.8717  & 94.6764 &4.9596 &-0.3073  &0.0157  &0.0154  &0.0154  &1400.00  &0.00032  &0.00032 &0.00032  \\ 
75\%              &255.0788  &167.0582 &5.0665 &-0.3055  &0.0398  &0.0402  &0.0377  &1400.00  &0.00047  &0.00047 &0.00047  \\ 
Max.              &427.4502  &396.3159 &5.2506 & 0.0    &0.3153  &0.4934  &0.2730  &1400.00  &0.24625  &0.24625 &0.24625  \\  
    \end{tabular}%
\end{ruledtabular}
\end{minipage}
\end{table*}

It is important to emphasize that in Qiskit~\citep{Qiskit}, qubits are arranged in little-endian order, where the qubit with the smallest index corresponds to the least significant bit. For instance, in a three-qubit system, it is represented as $|q_2q_1q_0\rangle$. Due to this convention, in practical implementations on IBM Quantum's hardware, the circuits displayed will appear horizontally flipped compared to their usual depiction.

\subsection{List of Abbreviations }

\begin{table}
    \centering 
    \begin{tabular}{ll}
    NISQ         &Noisy intermediate-scale quantum    \\
    GSA          &Grover search algorithm\\
    ${\cal SSO}$          &Squared statistical overlap\\ 
    ${\cal ASP}$          &Algorithm success probability\\
    ${\cal QST}$          &Quantum state tomography     \\
    $\Gamma$     &A set of an unstructured S elements\\
   $\pounds$    &The total number of possible answers\\
    StDev        & Standard deviation \\
    SE Mean      & The standard error of the mean \\
    $\mu$        & Population mean \\
    $\sigma^2$   & Population variance  \\
    CI           & confidence interval\\
    Eagle r3    & A quantum processor developed by IBM Quantum\\
    ${\cal F}_S$ &The state fidelity \\
    $q_i$        & The $i^{th}$  qubit \\
    $q_k$        &Ancillary qubit\\
$\ket{\chi^*}$                     &Single marked state\\
$\ket{\chi^*_1}$, $\ket{\chi^*_2}$ & Two marked states\\
$\ket{\psi_E}_\text{Single}$    & The expected state for a single search result\\
$\ket{\psi_E}_\text{Multi}$     & The expected state for a two-search results\\
    $T_1$         &Relaxation time    \\ 
    $T_2$         &Dephasing time    \\ 
    $\Delta $       &Qubit anharmonicity    \\ 
    $f_{\text{qubit}}$   &Qubit frequency    \\ 
    $\epsilon_{\text{readout}}$           &The qubit readout assignment error    \\ 
    $ P_{\text{m1p0}} $        &The qubit flip probabilities from $ \ket{0}$  to $ \ket{1}$    \\ 
    $ P_{\text{m0p1}} $        &The qubit flip probabilities from $ \ket{1}$  to $ \ket{0}$    \\ 
    $t_{\text{readout}}$       &Readout length    \\
    $\epsilon_{\text{ID}}$     &The error rate associated with the Identity gate  \\
    $\epsilon_{\text{SX}} $     &The error rate associated with the SX gate  \\
    $\epsilon_{\text{Pauli-X}}$     &The error rate associated with the Pauli-X gate  \\
    Re.             &Real part \\
    Im.             &Imaginary part \\
    $\rho$       &Density matrix \\
    $\mu$s          &Microseconds \\
    ns              &Nanoseconds \\
    GHz             &Gigahertz  \\
    MHz             &Megahertz  \\
    \end{tabular}
\end{table}

\begin{acknowledgments}

We acknowledge the use of IBM Quantum's superconducting quantum computers in this research. Nevertheless, the perspectives and findings  presented in this paper are solely those of the author and do not necessarily reflect the views of IBM Quantum. M. AbuGhanem extends special thanks to Prof. Diego Emilio Serrano (Panasonic Massachusetts Laboratory) and Frank Harkins (IBM Quantum developer advocacy) for providing assistance in questioning regarding retrieving extensive experimental results.

\end{acknowledgments}

\section*{Declarations}

\subsection*{Ethical Approval and Consent to participate}
Not applicable.

\subsection*{Consent for publication}
The author have approved the publication. This research did not involve any human, animal or other participants.

\subsection*{Availability of supporting data}
The datasets generated during and/or analyzed during the current study are included within this article.

\subsection*{Competing interests}
The author declares no competing interests.

\subsection*{Funding}

The author declares that no funding, grants, or other forms of support were received at any point throughout this research work.

\subsection*{Authors' contributions}

M. AbuGhanem: conceptualization, methodology, quantum programming, experimental implementations on IBM Quantum's quantum computers, data curation, formal analysis, statistical analysis, visualization, investigation, validation, writing, reviewing and editing. The author has approved the final manuscript.

\end{document}